\documentclass[aps,prl,superscriptaddress,twocolumn]{revtex4-2}

\usepackage[french,english]{babel}
\usepackage[utf8x]{inputenc}
\usepackage[T1]{fontenc}

\usepackage{times}

\usepackage{subfigure}
\usepackage{t1enc,latexsym,amsfonts, amssymb, amsmath,empheq}
\usepackage{bm}
\usepackage{graphics}
\usepackage{graphicx}
\usepackage{url}
\usepackage{soul}

\usepackage[usenames,dvipsnames,table]{xcolor}
\usepackage{color}
\definecolor{darkblue}{rgb}{0,0,0.6}
\definecolor{darkred}{rgb}{0,0,0.6}

\usepackage{hyperref}
\hypersetup{
    bookmarksopen=true,
    pdftitle="",
    pdfauthor="Agoritsas-Morse", 
    pdfsubject="",
    pdfstartview={FitH},
    pdfmenubar=true,
    colorlinks=true,
    urlcolor=darkblue,
    citecolor=darkblue,
   linkcolor=darkred
}

\newcommand{\argc}[1]{\left[#1\right]}
\newcommand{\arga}[1]{\left\lbrace #1\right\rbrace }
\newcommand{\argp}[1]{\left(#1\right)}

\newcommand{\horizontalline}{\begin{center} \rule{\textwidth}{0.5pt} \end{center}}

\begin{document}


\title{Memory formation in dense persistent active matter}

\author{Elisabeth Agoritsas}
\email[]{elisabeth.agoritsas@unige.ch}
\affiliation{Alphabetical order}
\affiliation{Department of Quantum Matter Physics (DQMP), University of Geneva, Quai Ernest-Ansermet 24, CH-1211 Geneva, Switzerland}

\author{Peter K. Morse}
\email[]{peter.k.morse@gmail.com}
\affiliation{Alphabetical order}
\affiliation{Department of Chemistry, Princeton University, Princeton, NJ 08544}
\affiliation{Department of Physics, Princeton University, Princeton, NJ 08544}
\affiliation{Princeton Institute of Materials, Princeton University, Princeton, NJ 08544}

\date{\today}

\begin{abstract}

Protocol-dependent states in structural glasses can encode a disordered, yet retrievable memory.
While training such materials is typically done via a \emph{global} drive, such as external shear, in dense active matter the driving is instead \emph{local} and spatio-temporally correlated.
Here we focus on the impact of such spatial correlation on memory formation.
We investigate the mechanical response of a dense amorphous packing of athermal particles,
subject to an oscillatory quasistatic driving with a tunable spatial correlation,
akin to the instantaneous driving pattern in active matter.
We find that the capacity to encode memory can be rendered comparable upon a proper rescaling on the spatial correlation,
whereas the efficiency in memory formation increases with motion cooperativity.

\end{abstract}

\maketitle

\paragraph{Introduction.}
\label{sec-intro}

Driven amorphous solids have recently shown great promise in understanding and characterizing material memory
\cite{keim_2019_RevModPhys91_035002,nagel_2023_JChemPhys158_210401}.
%
While structural disorder renders theoretical descriptions challenging \cite{arceri_landes_2020_Arxiv-2006.09725},
it can also be harnessed to engineer specific disorder-induced properties
\cite{kadic_2019_NatureReviewsPhysics1_198,yu_2021_NatureReviewsMaterials6_226,upadhyaya_amir_2018_PhysRevMaterials2_075201}.
%
From a memory perspective, specific states can be encoded---and retrieved---with carefully tailored driving protocols.
The challenge is then to characterize which states can be reached \cite{sastry_1998_Nature393_554,ozawa_2012_PhysRevLett109_205701,charbonneau_morse_2021_PhysRevLett126_088001,charbonneau_2023_PhysRevE108_054102},
and with how much training.
One class of such mechanical trainings are oscillatory athermal quasistatic shear,
either in particle-based \cite{kawasaki_berthier_2016_PhysRevE94_022615,leishangthem_2017_NatureComm8_14653,adhikari_sastry_2018_EPJE41_105,yeh_2020_PhysRevLett124_225502,bhaumik_2021_PNAS118_e2100227118}
%
or coarse-grained models \cite{khirallah_2021_PRL126_218005,kumar_2022_JChemPhys157_174504,liu_2022_JChemPhys156_104902,parley_2022_PhysRevLett128_198001}.
%
Here, a system is cyclically sheared starting from a given initial condition,
typically at equilibrium,
and may eventually reach a hysteretic limit cycle.
Depending on the maximum strain,
its mechanical response displays a dynamical yielding transition
from a regime where return-point memory is achievable
to a regime where plastic rearrangements prevent memory formation \cite{regev_2015_NatureComm6_8805}.

In parallel, a considerable effort is being deployed to understand dense active matter.
A central goal is to assess which phenomena from amorphous solids with \emph{passive} individual components are relevant, and thus exportable, to interacting active particles
\cite{janssen_2019_JPhysCondensMatter31_503002,berthier_2019_JChemPhys150_200901,divoux_2023_ArXiv-2312.14278}.
%
Characterizing the interplay between structural disorder and activity has become a flourishing field,
\textit{e.g.} to investigate the role mechanical signalling in confluent tissues \cite{
schoetz_2013_JRSocInterface10_20130726,
bi_2014_SoftMatter10_1885,
park_2015_NatMater14_1040,
bi_2015_NatPhys11_1074,
etournay_2015_eLife4_e07090,
bi_manning_2016_PhysRevX6_021011,
matoz_2017_SoftMatter13_3205,
matoz_agoritsas_2017_PhysRevLett118_158105,
merkel_2017_PhysRevE95_032401,
henkes_2020_NatureCommunications11_1405,
wang_2020_PNAS117_13541,
popovic_2021_NewJPhys23_033004,
divoux_2023_ArXiv-2312.14278}.
%
%
The key novelty is that  active systems are driven locally, at the scale of their individual components,
instead of globally \cite{marchetti_2013_RevModPhys85_1143}.
%
Nevertheless, active systems experience plasticity in sufficiently dense regimes,
affecting both their mechanical and rheological properties.
In light of this, we have established a direct link between 
global shear and spatially-correlated local driving \cite{morse_roy_agoritsas_2021_PNAS118_e2019909118},
akin to the instantaneous self-propulsion in active matter \cite{henkes_2020_NatureCommunications11_1405}.

In this Letter,
we focus on the role of such spatial correlation on memory formation.
Building on the connection
predicted and tested in \cite{morse_roy_agoritsas_2021_PNAS118_e2019909118,agoritsas_2021_JStatMech2021_033501},
we investigate the mechanical response of a dense amorphous packing of athermal particles,
subject to an oscillatory driving pattern with a tunable spatial correlation.
We focus on their stress-strain curves, which display a hysteretic behavior depending on the maximum strain and number of cycles.
Memory is quantified by comparing successive snapshots of configurations after each cycle.

Our study shows that by tuning the spatial correlation of the local driving, we can control both the capacity and efficiency to reach a limit cycle, and therefore to encode a memory.
Our findings suggest that increasing motion cooperativity is a generic strategy to enhance memory formation.


\begin{figure}[htbp]
\begin{center}
\includegraphics[width=\columnwidth]{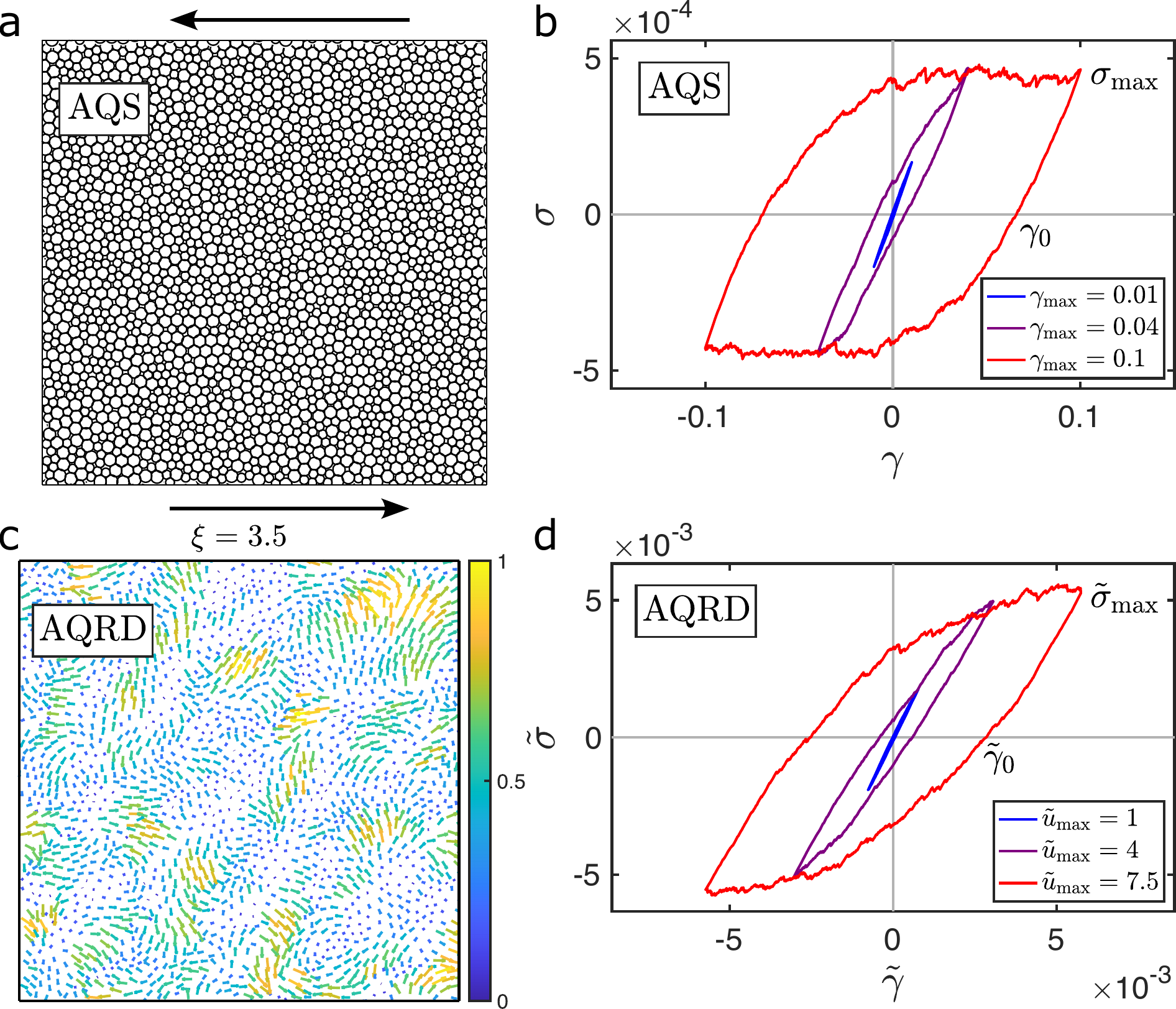}
\caption{
\textbf{Driving protocol and limit cycles}.
(a)~AQS illustration wherein
a global shear $\gamma$ is applied by Lees-Edwards (periodic) boundary conditions.
(b)~Averaged (apparent) limit cycles of stress-strain curves ${\sigma(\gamma)}$;
increasing the maximum strain $\gamma_{\mathrm{max}}$ unveils hysteretic behavior,
quantified by $\arga{\sigma_{\mathrm{max}},\gamma_0}$.
(c)~AQRD illustration wherein
local displacement vectors---akin to an instantaneous velocity pattern---have spatial correlation ${\xi=3.5}$ and periodic boundary conditions.
(d)~Corresponding averaged limit cycles of random stress-strain curves $\tilde{\sigma}(\tilde{\gamma})$;
increasing the maximum displacement amplitude $\tilde{u}_{\mathrm{max}}$ of individual particles (along their unnormalized displacement vectors) unveils hysteretic behavior,
quantified by $\arga{\tilde{\sigma}_{\mathrm{max}},\tilde{\gamma}_0}$.
(b)~and~(d) are measured on the last cycle (${t=10}$), and averaged over 30 different initial conditions.
}
\label{fig:stress-straincurves-compil}
\end{center}
\end{figure}

\paragraph{Model and driving protocol.}
\label{sec-model}
Simulations are performed on dense packings of
${N=2048}$
bidisperse Hertzian disks with a ${1:1.4}$ size ratio and ${1:1}$ number ratio,
under periodic boundary conditions in a square box of size $L$. 
The position of particle $i$ is given by $\mathbf{r}_i$.
We initialize the system at $\arga{\mathbf{r}_{i,0}}$ with the standard prescription of \cite{ohern_2003_PhysRevE680_11306}:
particles are uniformly distributed via a Poisson process (corresponding to an infinite temperature quench) at a packing fraction ${\varphi = 0.95}$---well above the jamming transition ${\varphi_J \approx 0.842}$~\cite{ohern_2002_PhysRevLett88_075507}---and minimized via the FIRE protocol~\cite{bitzek_2006_PhysRevLett97_170201} until the inherent state is found, while keeping the density fixed.
Distance units are chosen such that the smaller particles have unit diameter,
thus ${L \approx 50.1}$ and an average interparticle distance $\ell\approx1.1$.
The results of \cite{morse_roy_agoritsas_2021_PNAS118_e2019909118, morse_2020_PhysRevResearch2_023179,goodrich_2012_PhysRevLett109_095704,goodrich_2014_PhysRevE90_022138,dagois-bohy_2012_PhysRevLett109_095703} suggest that our choice of $\arga{\varphi,N}$ will quantitatively change certain scales, but not fundamentally alter the physics, provided that ${\varphi > \varphi_J}$.

We consider two protocols
illustrated in Fig.~\ref{fig:stress-straincurves-compil}:
a \textit{global} athermal quasistatic shear (AQS)~\cite{maloney_lemaitre_2006_PhysRevE74_016118}
and a \textit{local} athermal quasistatic random displacement (AQRD)~\cite{morse_roy_agoritsas_2021_PNAS118_e2019909118}.
Full details of each implementation are given in the above references.
For AQS, we use Lees-Edwards boundary conditions to apply a dimensionless strain increment $\Delta\gamma=10^{-4}$, minimize via FIRE, and measure the corresponding stress using the Born-Huang approximation~\cite{book_born_dynamical_1954}.
The total accumulated strain is denoted $\gamma$,
and the corresponding shear stress $\sigma$.
For AQRD,
we assign to each particle $i$ a local displacement vector $\mathbf{c}_i$ that we keep constant.
We label the tunable driving pattern $|c\rangle = \arga{\mathbf{c}_i}$, with the ket notation indicating an $Nd$-dimensional vector.
Throughout this work,
we generate a Gaussian field $\mathcal{C}(\mathbf{r})$ of correlation length $\xi$
as in \cite{morse_roy_agoritsas_2021_PNAS118_e2019909118},
and combine it with the initial configuration  ${\arga{\mathbf{c}_i \equiv \mathcal{C}(\mathbf{r}_{i,0})}}$.
This is akin to freezing the instantaneous velocity field in dense active matter, or alternatively working at timescales smaller than the persistence time.
A displacement increment $\Delta\tilde{u}=10^{-4}L$ is then applied along $|c\rangle$,
and a constrained minimization (via FIRE) is performed using a Lagrange multiplier along $|c\rangle$,
globally prohibiting motion along $|c\rangle$.
The total accumulated AQRD displacement is denoted $\tilde{u}$.
For AQRD, the relevant random `strain' and `stress' are defined, respectively, as
\begin{equation}
\tilde{\gamma} \equiv \frac{\tilde{u}}{L \sqrt{N/3}},
\quad 
\tilde{\sigma} \equiv \frac{1}{L^2}\frac{\partial U}{\partial \tilde{\gamma}} = -\langle F|c\rangle \frac{1}{L}\sqrt{\frac{N}{3}},
\label{eq:gammaTilde-sigmaTilde-Def}
\end{equation}
where $|F\rangle$ is the $Nd$ dimensional residual force.
Between plastic rearrangements, an apparent elastic modulus $\tilde{\mu}$ can be defined  as ${\Delta \tilde{\sigma}=\tilde{\mu} \, \Delta \tilde{\gamma}}$.

In both AQS and AQRD,
we apply an oscillatory strain such that strain/displacement is incrementally increased to $\arga{\gamma_{\mathrm{max}},\tilde{u}_{\mathrm{max}}}$,
then similarly decreased to $\arga{-\gamma_{\mathrm{max}},-\tilde{u}_{\mathrm{max}}}$,
and the process is repeated.
Each full cycle is labelled by the integer $t$.
We consider the same 30 initial seed configurations, which fix both ${\arga{\mathbf{r}_{i,0}}}$ and the initial driving field $\mathcal{C}(\mathbf{r})$.
We record the corresponding stress-strain curves ${\sigma(\gamma)}$ and ${\tilde{\sigma}(\tilde{\gamma})}$, along with successive snapshots of the system at maximum strains.
The simulations thus depend on
the maximum driving amplitude $\arga{\gamma_{\mathrm{max}},\tilde{u}_{\mathrm{max}}}$,
the total number of cycles $t$,
and the spatial correlation $\xi$ for AQRD.
Typical mechanical responses are illustrated in Fig.~\ref{fig:stress-straincurves-compil} at our last cycle ${t=10}$.
See SI for more details on the numerics, the range of explored parameters,
and examples of individual trajectories.

\begin{figure}[bp]
\begin{center}
\includegraphics[width=\columnwidth]{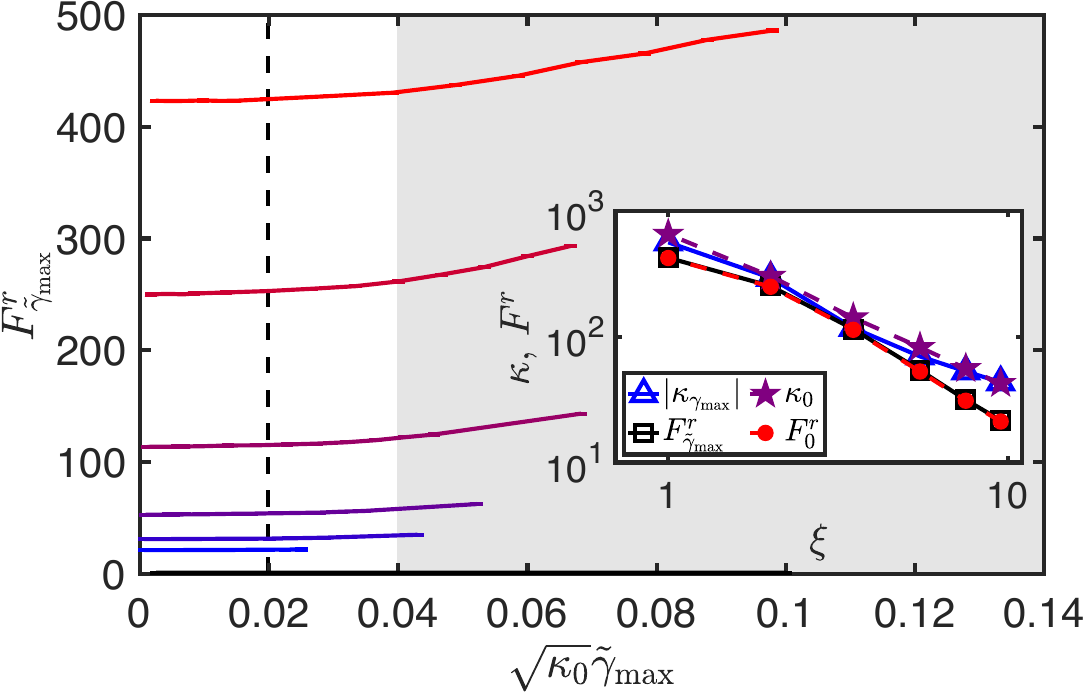}
\caption{\textbf{Driving patterns vs elastic modulii at the limit cycles}.
\textit{Main}:~Variance of the local driving at the maximum strain, after ${t=10}$ cycles averaged over $30$ initial configurations,
normalized by the reference value $\mathfrak{F}_0^{\mathrm{AQS}}/\ell^2=3.2\times10^{-6}$;
with the notation ${F}^r \equiv \mathfrak{F}/\mathfrak{F}_0^{\mathrm{AQS}}$.
\textit{Inset}:~
$\xi$-dependent decreasing trend of $F^r_0$ and $\kappa_0$ measured initially,
and $F^r_{\tilde{\gamma}_\mathrm{max}}$ and $\kappa_{\tilde{\gamma}_\mathrm{max}}$ measured at a small effective strain $\sqrt{\kappa_0}\tilde{\gamma}_\mathrm{max} \approx 0.02$ (dashed line in main).
}
\label{fig:pattern-charact-elastic-modulii}
\end{center}
\end{figure}

\begin{figure*}[htbp]
\begin{center}
\includegraphics[width=\linewidth]{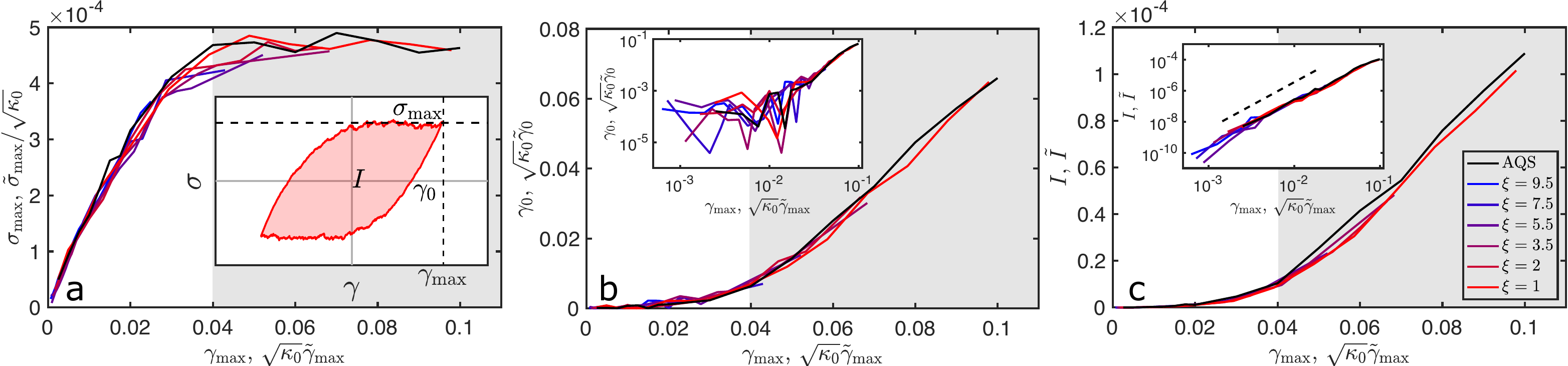}
\caption{
\textbf{Geometrical features of hysteretic limit cycles}.
When scaled appropriately with $\kappa_0(\xi)$,
the averaged stress-strain loops of Fig.~\ref{fig:stress-straincurves-compil} (after ${t=10}$ cycles)
collapse onto master curves for AQS (black) and AQRD (red through blue for increasing $\xi$).
(a)~The maximal stress crosses over from a linear behavior in the `pre-yielding' regime ($\sqrt{\kappa_0} \tilde{\gamma}_\mathrm{max} \lesssim
 0.04$), to a plateau above this range (light-grey shaded).
(b)~For the entire range considered,
${\gamma_0 \propto \gamma_\mathrm{max}^2}$ and ${\tilde{\gamma}_0 \propto \tilde{\gamma}_\mathrm{max}^2}$,
though this metric is sensitive to individual rearrangements and thus noisy for small effective strains (as shown in the log-scaled inset).
(c)~The surface spanned by the hysteretic loop
($I\equiv\int \sigma d\gamma$, $\tilde{I}\equiv\int \tilde{\sigma}d\tilde{\gamma}$)
collapses for the entire range.
Given (a)-(b),
this proxy for the energy density dissipated in one limit cycle displays a cubic behavior
${I\sim \gamma_{\mathrm{max}}^3}$
and ${\tilde{I} \sim \tilde{\gamma}_{\mathrm{max}}^3}$
in the pre-yielding regime (dashed line in the log scaled inset).
}
\label{fig:averaged-limitcycles}
\end{center}
\end{figure*}

\paragraph{Driving pattern characterization vs average elastic modulus.}
\label{sec-effective-pattern-characterization-elastic-modulus}
The analytical predictions in \cite{morse_roy_agoritsas_2021_PNAS118_e2019909118,agoritsas_2021_JStatMech2021_033501} suggest a statistical equivalence between AQS-AQRD stress-strain curves,
provided that we compare
the AQS strain $\gamma$ to the effective random strain ${\tilde{\gamma}_{\text{eff}}=\sqrt{\kappa} \tilde{\gamma}}$,
and the AQS shear stress $\sigma$ to ${\tilde{\sigma}_{\text{eff}}=\tilde{\sigma}/\sqrt{\kappa} }$.
The rescaling factor $\kappa(\xi)$ is measured as the ratio of the average apparent modulus $\tilde{\mu}(\xi)$  (of the driving pattern $\vert c \rangle$),
to its counterpart $\mu_{\text{AQS}}$ (under global shear).
This ratio depends on the disorder of the driving pattern, decreasing with increased cooperativity (\textit{i.e.}~larger $\xi$).
For pairwise-interacting particles,
this is quantified by  the variance of relative displacements on interacting pairs
${\mathbf{c}_{ij}= \mathbf{c}_i - \mathbf{c}_j}$
nondimensionalized with $\ell$, denoted ${\mathfrak{F} (\xi)/\ell^2}$.
The analytical prediction further states that ${\kappa (\xi) \propto \mathfrak{F} (\xi)/\ell^2}$,
providing a direct and straightforward link between the mechanical response of the dense packing and the statistical features of the driving pattern \cite{morse_roy_agoritsas_2021_PNAS118_e2019909118,agoritsas_2021_JStatMech2021_033501}.

Both $\kappa$ and $\mathfrak{F}$ are physical quantities measurable in an agnostic way from a given snapshot of the system (see SI).
We denote
their initial values $\arga{\kappa_0,\mathfrak{F}_0}$, and at the maximum strain in the last cycle $\arga{\kappa(\tilde{\gamma}_{\mathrm{max}}),\mathfrak{F} (\tilde{\gamma}_{\mathrm{max}})}$, presumably in the limit cycle, if it exists.
We report their values in Fig.~\ref{fig:pattern-charact-elastic-modulii} for different driving correlation $\xi$.
To make these values comparable, we rescale the random strain with the initial $\kappa_0(\xi)$, so as not to depend on the subsequent driving protocol.
We see that ${\mathfrak{F} (\tilde{\gamma}_{\mathrm{max}})}$ is constant 
for ${\sqrt{\kappa_0} \tilde{\gamma}_{\mathrm{max}} \lesssim 0.04}$, which is roughly the yielding point as discussed later (see Fig.~\ref{fig:averaged-limitcycles}).
The variance $\mathfrak{F}$ significantly increases beyond that, as can be seen for the three smallest-$\xi$ curves,
showing that the driving pattern has become more disordered (\textit{i.e.} effectively a smaller $\xi$) upon cycling; and above ${\sqrt{\kappa_0} \tilde{\gamma}_{\mathrm{max}} \lesssim 0.04}$,
we later show that a limit cycle is not reached \textit{per se}.
In the inset, we show the $\xi$-dependence
at a small effective strain ${\sqrt{\kappa_0} \tilde{\gamma}_{\max} \approx 0.02}$,
by selecting maximum effective strains carefully for each $\xi$.
As expected, $\kappa$ follows the same decreasing trend as the pattern variance $\mathfrak{F}$;
physically, it is easier to deform a material with larger motion cooperativity because it has a lower apparent elastic modulus.

A fixed driving pattern is implicitly enforced in our analytical predictions \cite{morse_roy_agoritsas_2021_PNAS118_e2019909118,agoritsas_2021_JStatMech2021_033501},
as particle displacements are then restricted to a distance $\sim 1/\sqrt{d}$ in the limit of dimension ${d \to \infty}$.
Note that we restricted ourselves to AQRD protocols with ${\tilde{u}_{\mathrm{max}}/N \ll 1 \leq \xi}$, so that this initial pattern mostly survives upon successive cycles, despite the plastic rearrangements occurring until (and at) the limit cycles.
This guarantees that our driving protocol keeps the system within the validity range of the infinite-dimensional limit predictions.
The overall persistence of the initial pattern is confirmed by
${\mathfrak{F}(\tilde{\gamma}_{\mathrm{max}})\approx \mathfrak{F}_0}$
and
${\kappa(\tilde{\gamma}_{\mathrm{max}})\approx \kappa_0}$,
further supporting our choice of $\kappa_0(\xi)$ to rescale all subsequent plots.

\paragraph{Collapsing the averaged limit cycles.}
\label{sec-averaged-limitcycles-rescaling}
We can now use this rescaling to compare the stress-strain curves as plotted in Fig.~\ref{fig:stress-straincurves-compil}.
The relevance of the infinite-dimensional analytical predictions for our model was successfully tested in \cite{morse_roy_agoritsas_2021_PNAS118_e2019909118}
but only in the pre-yielding regime starting from $\arga{\mathbf{r}_{i,0}}$,
\textit{i.e.}~in the very first cycle of our oscillatory protocol.
Here we use it as a tool to render comparable
the geometrical features of mechanical hysteresis, as defined and plotted in Fig.~\ref{fig:averaged-limitcycles}:
$\lbrace \sigma_\mathrm{max},\gamma_0,I \rbrace$ for AQS,
$\lbrace {\tilde{\sigma}_\mathrm{max},\tilde{\gamma}_0,\tilde{I}}\rbrace$ for AQRD, where $I\equiv\int \sigma d\gamma$ and $\tilde{I}\equiv\int \tilde{\sigma}d\tilde{\gamma}$.
These quantities are measured and averaged on the last cycle as in Fig.~\ref{fig:stress-straincurves-compil}, presumably at the limit cycle (when reachable).
As predicted analytically \cite{agoritsas_2021_JStatMech2021_033501},
we find that they can be collapsed onto master curves using the initial ratio $\kappa_0(\xi)$.
Thus
the mechanical response for a given local displacement $\tilde{u}_{\mathrm{max}}$
under oscillatory AQRD
can be pushed either inside or outside an effective `pre-yielding' regime, simply by tuning the spatial correlation $\xi$ of the driving pattern.
This explains why the simulation range is effectively compressed in Fig.~\ref{fig:averaged-limitcycles} when increasing $\xi$.
On the other hand, increasing cooperativity promotes an increased efficiency
for a given strain,
\textit{i.e.}~less dissipated energy per cycle $\propto I,\tilde{I}$.


\begin{figure*}[htp]
\begin{center}
\includegraphics[width=\linewidth]{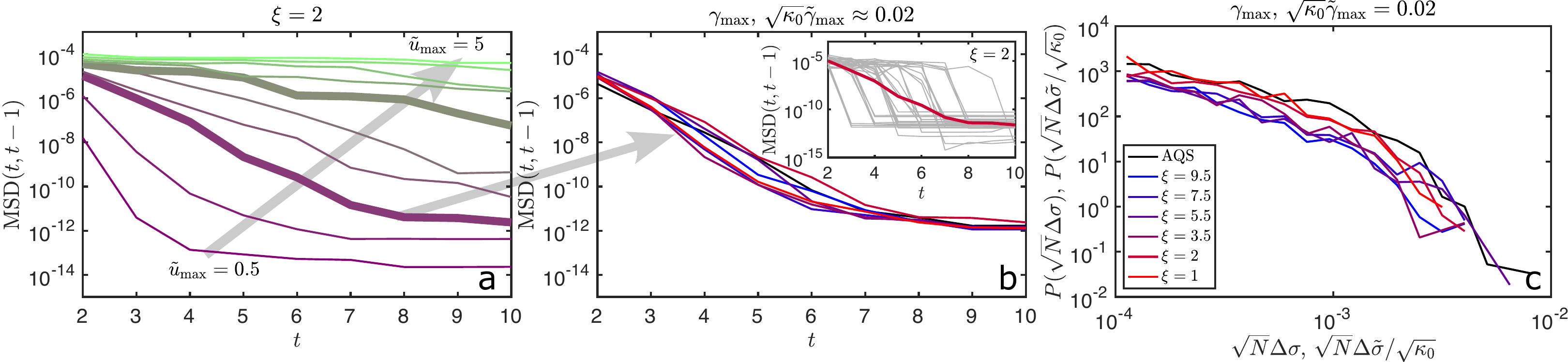}
\caption{\textbf{Quantifying memory formation, and statistical features towards the limit cycle.}
(a)~MSD function at fixed spatial correlation averaged over 30 initial configurations with increasing maximum strain
(thick purple line ${\sqrt{\kappa_0}\tilde{\gamma}_{\mathrm{max}} \approx 0.02}$,
green $\approx 0.04$).
(b)~Collapse of MSD functions for different $\xi$ and AQS, at comparable small effective strain in the pre-yielding regime
(same dataset as in Fig.~\ref{fig:pattern-charact-elastic-modulii} inset, see also SI);
the inset shows the wide spread of individual configurations and their averaged curve using the geometric mean.
(c)~Collapse of the corresponding stress-drop distributions on the first linear regime of the last cycle (${t=10}$), using the same rescaling as in Fig.~\ref{fig:averaged-limitcycles}.
}
\label{fig-memory}
\end{center}
\end{figure*}

\paragraph{Memory characterization.}
\label{section-memory-characterization}

We now examine the implications of these findings for memory formation.
We quantify the convergence to a limit cycle
by comparing successive stroboscopic snapshots at the maximum strain at each cycle $t$, with the mean-square-displacement (MSD) of the difference,
${\text{MSD}(t,t-1) =
\left\langle \frac{1}{N} \sum_i |\mathbf{r}_i(t) - \mathbf{r}_i(t-1)|^2 \right\rangle}$
where brackets denote the average over initial configurations.
By increasing the maximum strain we cross over from a regime with an exponential decay to a limit cycle
to a regime where the existence of a limit cycle is not warranted [Fig.~\ref{fig-memory}(a)].
This crossover occurs at the `yielding point' already evidenced in Fig.~\ref{fig:averaged-limitcycles}(a).
At sufficiently large $\tilde{\gamma}_{\mathrm{max}}$,
plastic rearrangements prevent the system from locking into a limit cycle, yielding a finite MSD plateau when comparing configurations stroboscopically.
This threshold can be tuned with the motion cooperativity $\xi$ [Fig.~\ref{fig:averaged-limitcycles}]:
Fig.~\ref{fig-memory}(b) shows how the same $\kappa_0$ rescaling can further be used to collapse the averaged memory decay upon AQRD cycling.
Put simply, a limit cycle (memory) can be achieved in a similar way
(\textit{e.g.} same number of cycles) by jointly tuning $\tilde{\gamma}_\mathrm{max}$ and $\xi$ so as to keep the effective strain fixed.
In practice, when increasing the motion cooperativity, a smaller absolute strain is typically required to achieve the same memory training.

\paragraph{Mean-field statistical features in the limit cycles.}
\label{sec-MF-features-pdf}

In \cite{morse_roy_agoritsas_2021_PNAS118_e2019909118}, the link between AQS and AQRD was shown
on three mean-field metrics of the disordered landscape explored by a typical glassy system---via the collapse of the elastic-modulus, the stress-drops and strain-steps distributions---in the very first cycle.
Fig.~\ref{fig-memory}(c) shows similarly a good collapse of the stress-drop distributions at the last cycle, for the same composite dataset at small effective strain.
This then supports the claim that the predicted scaling holds
for the statistical features of the landscape visited along the \emph{whole} AQRD/AQS training.

\paragraph{Conclusion.}
\label{section-conclusion}
We examined the role of motion cooperativity in memory formation,
focusing on oscillatory driving protocols under spatially-correlated local driving.
We found that the hysteretic mechanical response could be rendered comparable by rescaling the stress-strain curves with $\kappa(\xi) \propto \mathfrak{F}(\xi)$, as predicted 
in \cite{morse_roy_agoritsas_2021_PNAS118_e2019909118,agoritsas_2021_JStatMech2021_033501},
and that this connection applies to memory dynamics as well.
Increasing the driving spatial correlation pushes the system towards the pre-yielding regime where memory is achievable,
\textit{(i)}~with a more efficient training (\textit{i.e.}~less energy dissipated per cycle)
and \textit{(ii)}~requiring lower maximum strain to reach a given target state. 

Conversely, more localized fields open the possibility to encode multiple localized memories into the same sample,
of great interest for materials engineering.


In genuine dense active matter,
the velocity field can be self-generated \cite{henkes_2020_NatureCommunications11_1405} or externally tuned, for instance, by chemical signalling in biological tissues \cite{li_2021_PNAS118_e2025717118} or remote-sync in crowds.
Our findings suggest that allowing for more cooperativity is a good strategy to robustly and efficiently encode  memory in a structurally disordered system.
Here we focused on the role of spatial correlation, but its interplay with a finite persistence time remains an open issue \cite{behera_2023_PhysRevX13_041043,goswami_2023_Arxiv-2312.01459},
as reinstating time also implies moving beyond the quasistatic limit.
For extremely persistent active systems and small driving amplitude, the protocol of \cite{keta_2022_PhysRevLett129_048002,keta_2023_SoftMatter19_3871} could be adapted to include a finite spatial correlation in a quasistatic-like settings.
Hence we speculate that our findings extend at low driving rate---given precautions to stay within the validity range of these predictions---albeit with additional challenges to rationalize memory formation at finite rates \cite{agoritsas_bares_2023_Arxiv-2308.05603}.

At a more abstract level,
our oscillatory AQRD protocol
allows us to tune the energy landscape effectively explored by the system.
The exact AQS/AQRD equivalence in infinite dimension \cite{agoritsas_2021_JStatMech2021_033501}
extends to the mean-field statistical features of this landscape.
A corollary issue is thus to understand how its complex structure is spatially connected---casting it for instance on random transition graphs between mechanically stable configurations as for AQS \cite{mungan_witten_2019_PhysRevE99_052132,mungan_2019_PhysRevLett123_178002,regev_2021_PhysRevE103_062614,mungan_2021_PhysRevLett127_248002}---using our imposed cooperativity as an artificial probe of the underlying graph topology.
This would be key to understanding which types of memory can be encoded in dense active matter.


\begin{acknowledgments}

E.A. acknowledges partial support from the Swiss National Science Foundation by the SNSF Ambizione Grant No.~PZ00P2{\_}173962, and from the Simons Foundation (Grant No.~348126 to S.~Nagel).
The simulations were developed at Princeton University on the Adroit cluster and performed at the University of Geneva on the Mafalda cluster.

\end{acknowledgments}





\cleardoublepage
\newpage
\onecolumngrid
\appendix

\horizontalline
\begin{center}
\textsc{Supplementary information}
\end{center}
\horizontalline

\vline

We provide supplementary information on the following items:
\begin{enumerate}
\item Details on the numerical recipe and list of parameters
\item Individual and averaged stress-strain curves
\item Driving pattern characterization vs average elastic modulus
\item Geometrical features of hysteretic limit cycles
\item Quantifying amnesia
\end{enumerate}

\horizontalline


\section{Details on the numerical recipe and list of parameters}
\label{sec-numerical-recipe-parameters}

\paragraph{Interaction potential.}
As in Ref.~\cite{morse_roy_agoritsas_2021_PNAS118_e2019909118}, we perform simulations of dense packings of bidisperse soft disks in dimension $d=2$ with a Hertzian pair potential,
wherein the position of particle $i$ with radius $\rho_i$ is given by $\mathbf{r}_i$, and the inter-particle distance of a pair $(i,j)$ by $r_{ij}=|\mathbf{r}_i - \mathbf{r}_j|$.
Using a dimensionless overlap ${\varepsilon_{ij}=1-r_{ij}/(\rho_i+\rho_j)}$, the total energy of the configuration is given by
${U = \frac{1}{\alpha}\sum_{ij}\Theta(\varepsilon_{ij})\varepsilon_{ij}^\alpha}$,
where $\Theta$ is the Heaviside function, $\alpha=2.5$ for Hertzian particles, and the sum is over all unique pairs of particles. This is considered a standard system~\cite{ohern_2003_PhysRevE680_11306}, as it allows the exploration of packings with densities above the jamming transition. Here, overlaps between particles are allowed. While the analysis of these systems would be similar under a Hookean potential ($\alpha=2$), a Hessian analysis is more generalizable when $\alpha \neq 2$. We do not expect different potentials to yield drastically different results, though stress and strain scales will of course differ with different interactions.

\vline

\paragraph{System size and packing fraction.}
We consider packings of ${N=2048}$ particles,
under periodic boundary conditions in a square box with side lengths ${L_x=L_y}$.
In order to ensure a disordered system, we use a number ratio of ${50:50}$ and a size ratio of ${1:1.4}$, known to prevent crystallization \cite{ohern_2002_PhysRevLett88_075507, ohern_2003_PhysRevE680_11306}.
Distance units are chosen such that the smaller particles have unit diameter.
We initialize the system at $\arga{\mathbf{r}_{i,0}}$ with the standard prescription of Ref.~\cite{ohern_2003_PhysRevE680_11306}:
particles are uniformly distributed via a Poisson process (corresponding to an infinite temperature quench) at a packing fraction ${\varphi = 0.95}$---well above the jamming transition ${\varphi_J \approx 0.842}$~\cite{ohern_2002_PhysRevLett88_075507}---and minimized via the FIRE protocol~\cite{bitzek_2006_PhysRevLett97_170201} until the inherent state is found.
(Note that while the packing fraction changes during a hard-sphere crunch, it remains fixed during soft-sphere minimizations such as ours.)
Fixing the packing fraction thus sets the system size to
$L_x=L_y=\sqrt{N\pi(1.4^2+1)/8\varphi}\approx 50.1$.
The average interparticle distance $\ell$ can be computed from the number density $\rho\equiv N/(L_xL_y)$, as $\ell=\rho^{-1/2}\approx1.1$,
signalling significant overlaps given the bidispersity.
In Ref.~\cite{morse_roy_agoritsas_2021_PNAS118_e2019909118} we report finite-size scalings for smaller packings with ${N \in \arga{64,128,256,512,1024}}$ and notice qualitatively similar results for all sizes considered.
Note however that we considered in Ref.~\cite{morse_roy_agoritsas_2021_PNAS118_e2019909118} a slightly lower packing fraction ${\varphi=0.94}$, also well above the jamming point. There we also considered a range of pressures (and thus a range of $\varphi$) obvserving qualitatively similar behavior providing that $\varphi>\varphi_J$. We choose packing fractions well above $\varphi_J$ to minimize the number of rattlers, which add numerical complications to memory formation.

\vline

\paragraph{Athermal quasistatic shear (AQS) driving protocol.}
Full details are given in
Refs.~~\cite{maloney_lemaitre_2006_PhysRevE74_016118, morse_2020_PhysRevResearch2_023179}.
We use Lees-Edwards boundary conditions to apply a dimensionless strain increment $\Delta\gamma=10^{-4}$, minimize via FIRE, and measure the corresponding stress using the Born-Huang approximation~\cite{book_born_dynamical_1954} as $\sigma=-\frac{1}{L_xL_y}\sum_{ij}r_{ij}^xf_{ij}^y$, where $r_{ij}^x$ denotes the $x$-component of the distance between contacting particles and $f_{ij}^y$ is the $y$-component of the force between them.
The total accumulated strain is denoted $\gamma$. 

\vline

\paragraph{Athermal quasistatic random-displacement (AQRD) driving protocol.}
Full details are given in Ref.~\cite{morse_roy_agoritsas_2021_PNAS118_e2019909118}.
We assign each particle $i$ a local displacement vector $\mathbf{c}_i$ that is kept constant, indicating persistant motion.
We denote $|c\rangle = \arga{\mathbf{c}_i}$ as a tunable driving pattern, with the ket notation indicating an $Nd$-dimensional vector.
For each initial configuration,
we generate a random Gaussian-correlated field $\mathcal{C}_{\xi}(\mathbf{r})$
as in the Supplementary information of Ref.~\cite{morse_roy_agoritsas_2021_PNAS118_e2019909118}.
The field is Gaussian distributed with zero mean ${\langle \mathcal{C}_{\xi}(\mathbf{r}) \rangle=0}$
and a two-point correlator
${\langle \mathcal{C}_{\xi}(\mathbf{r}) \mathcal{C}_{\xi}(\mathbf{r'}) \rangle=f(\vert{\mathbf{r}-\mathbf{r}'})}$,
where $f(r)$ is a Gaussian function of standard deviation which is proportional to $\xi$.
Moreover, $|c\rangle$ is specifically engineered to be compatible with the periodic boundary conditions;
global shear then corresponds to having a correlation length $\xi$ which is twice the size of the box \cite{morse_roy_agoritsas_2021_PNAS118_e2019909118}.
This random field is then combined with the initial configuration  ${\arga{\mathbf{c}_i \equiv \mathcal{C}_{\xi}(\mathbf{r}_{i,0})}}$, and a normalization is chosen such that $\langle c | c \rangle = 1$.
This process is akin to freezing the instantaneous velocity field in dense active matter, or alternatively working at timescales smaller than the persistence time of the activity.
A small displacement increment $\Delta\tilde{u}=10^{-4} L$ is then applied along $|c\rangle$ and a constrained minimization (via FIRE) is performed using a Lagrange multiplier along $|c\rangle$, such that motion along $|c\rangle$ is globally prohibited.
The total accumulated AQRD displacement is denoted $\tilde{u}$.
For AQRD, the relevant strain and stress are then defined, respectively:
\begin{equation}
\tilde{\gamma} \equiv \frac{\tilde{u}}{L_y\sqrt{N/3}},
\quad 
\tilde{\sigma} \equiv \frac{1}{L_xL_y}\frac{\partial U}{\partial \tilde{\gamma}} = -\langle F|c\rangle \frac{1}{L_x}\sqrt{\frac{N}{3}},
\label{eq:gammaTilde-sigmaTilde-Def}
\end{equation}
where $|F\rangle$ is the $Nd$ dimensional residual force.
In Ref.~\cite{morse_roy_agoritsas_2021_PNAS118_e2019909118} the factors of $3$ in Eq.~\eqref{eq:gammaTilde-sigmaTilde-Def} were mistakenly listed as factors of $12$ due to an incorrect choice of convention for the representation of the Lees-Edwards shear.
The argument comes from the (correct) statement that the AQS field for a small strain $\gamma$ on particle $i$ gives a displacement $\vec{u}_i = (y_i-L_y)\hat{x}$.
If, however, we quantify the total linear displacement of particles this expression erroneously yields $|u(\gamma)| = \gamma[\sum_i(y_i-L_y)^2]^{1/2}\approx \gamma L_y\sqrt{N/12}$ assuming a uniform distribution of $N$ particles.
This form biases towards small displacements because it uses an absolute value of a quantity centered on $0$.
Instead, a parameterization using $\vec{u}_i = y_i\hat{x}$ explicitly allows only positive values of displacement within the fundamental cell and thus offers an unbiased estimator when calculating total displacement.
We then have
\begin{equation}
|u(\gamma)| = \gamma\big[\sum_i y_i^2\big]^{1/2} \approx \gamma L_y\sqrt{\frac{N}{3}}
\end{equation}
when assuming a uniform distribution of $N$ particles in the box.
When calculating actual displacements, one must be careful at the top and bottom edges, however, to use image particles such that $\vec{u}$ is continuous.
Note that the AQRD pattern $\mathcal{C}(\mathbf{r})$ is specifically engineered to have a perfect periodicity in standard square boxes, so issues at the edge do not arise.

\vline

\paragraph{Oscillatory driving protocol and averaged mechanical response.}
In both AQS and AQRD,
we apply an oscillatory strain such that strain/displacement is incrementally increased to $\arga{\gamma_{\mathrm{max}},\tilde{u}_{\mathrm{max}}}$
then similarly decreased to $\arga{-\gamma_{\mathrm{max}},-\tilde{u}_{\mathrm{max}}}$,
and the process is repeated.
Each full cycle is labelled by the integer $t$, and we cycle up to ${t=10}$.
We consider the same 30 initial seed configurations, which fix both ${\arga{\mathbf{r}_{i,0}}}$ and the initial driving field $\mathcal{C}_{\xi}(\mathbf{r})$.
We record the corresponding stress-strain curves ${\sigma(\gamma)}$ and ${\tilde{\sigma}(\tilde{\gamma})}$, along with successive snapshots of the system at maximum and intermediate strains.
The simulations thus depend on
the maximum driving amplitude $\arga{\gamma_{\mathrm{max}},\tilde{u}_{\mathrm{max}}}$,
the total number of cycles $t$,
and the spatial correlation $\xi$ for AQRD.

\vline

\paragraph{List of parameters.}
We chose to explore, in an agnostic way, a similar range for the global strain $\gamma_{\mathrm{max}}$ and the local displacement amplitude $\tilde{u}_{\mathrm{max}}$,
with small increments in steps $10^{-4}$ in AQS and $10^{-4} L$ in AQRD.
This means that we need a table of conversion to go
first from $\tilde{u}_{\mathrm{max}}$ to the associated random strain $\tilde{\gamma}_{\mathrm{max}}$ using Eq.~\eqref{eq:gammaTilde-sigmaTilde-Def} (with ${N=2048}$ and ${L_x=50.1}$, imposed by $\varphi=0.95$),
and secondly to the effective strain ${\sqrt{\kappa(\xi)} \tilde{\gamma}_{\mathrm{max}}}$ choosing to use the initial $\kappa_0$.
The `x's indicate data points that were not run, and are therefore unavailable.
For the Figures displaying ${\sqrt{\kappa_0} \tilde{\gamma}_{\mathrm{max}}\approx 0.02}$, we use the values highlighted in yellow in Table~\ref{tab:list-parameters}.
Note that even the largest amplitudes ($\gamma_\mathrm{max}=0.1$ for AQS
and 
$\tilde{u}_\mathrm{max}/L=0.15$ for AQRD) yield average particle displacements $|u_\mathrm{max}|/N\approx0.064$ and $|\tilde{u}_\mathrm{max}|/N\approx0.0037$ in units of the smaller particle diameter, which are well below $1$, and thus also below the average interparticle distance $\ell \approx 1.1$.
This is self-consistent with the naive validity range of the infinite-dimensional formulation.
Note that these calculations assume linear response, but rearrangements will increase the size of displacements significantly. For this reason, we do not push to larger strains.
We use the following strain values (noting, however, that $\tilde{u}_\mathrm{max}/L = 0.15$ is omitted for $\xi\in \arga{1,2,9.5}$):

\begin{itemize}

\vspace{-2pt}
\item For AQS:\\ ${\gamma_{\mathrm{max}} \in \arga{0.0025, 0.0050, 0.0075, 0.01, 0.0125, 0.0150, 0.0175, 0.02, 0.03, 0.04, 0.05, 0.06, 0.07, 0.08, 0.09, 0.1}}$

\vspace{-2pt}
\item For AQRD at ${\xi \in \arga{1.0, 2.0, 3.5, 5.5, 7.5, 9.5}}$:\\
${\tilde{u}_{\mathrm{max}}/L \in \arga{0.0025, 0.0050, 0.0075, 0.01, 0.0125, 0.0150, 0.02, 0.03, 0.04, 0.05, 0.06, 0.07, 0.08, 0.09, 0.1, 0.15}}$,
\end{itemize}

\begin{table}[htbp]
\centering
    \begin{tabular}{|c|l|c|c|c|c|c|c|c|} \hline 
           &&  $\xi$&  1&  2.0&  3.5&  5.5&  7.5&  9.5\\ \hline 
           &&  $\kappa_0(\xi)$ &  650(40)&  300(20)&  142(11)&  82(6)&  56(4)&  43(4)\\\hline \hline \hline
          $\tilde{u}_{\mathrm{max}}/L$ &$\tilde{u}_{\mathrm{max}}$ &  $\tilde{\gamma}_{\mathrm{max}}$ $[10^{-3}]$&  \multicolumn{6}{|c|}{$\sqrt{\kappa_0} \tilde{\gamma}_{\mathrm{max}}$ $[10^{-2}]$}\\ \hline 
          0.0025 & 0.1251 &  0.096 & 0.244(8)& 0.167(6)& 0.114(4)& 0.087(3)& 0.072(3) &  0.062(3)\\ \hline 
          0.0050 &0.2502&  0.191 & 0.488(16)& 0.333(12)& 0.228(9)& 0.174(6)& 0.143(6)  &  0.125(6)\\ \hline 
          0.0075 & 0.3754 &  0.287  & 0.73(2)& 0.50(18)& 0.342(13)& 0.261(10)& 0.215(9) &  0.187(9)\\ \hline 
          0.01 & 0.5006  &  0.383 & \cellcolor{cyan}0.98(3)& 0.67(2)& 0.455(18)& 0.348(13)& 0.286(11) &  0.250(11)\\ \hline 
          0.0125 & 0.6257 &  0.478 & 1.22(4)& 0.83(3)& 0.57(2)& 0.434(16)& 0.358(14) &  0.312(15) \\ \hline 
          0.0150 & 0.7509 &  0.574 & 1.46(5)& \cellcolor{cyan}1.00(4)& 0.68(3)& 0.521(19)& 0.429(17) &  0.375(18)\\ \hline 
 0.0175 & 0.8760 & 0.670 & x& x& x& x& x& x\\ \hline 
 0.02 & 1.001 & 0.765 & \cellcolor{yellow}1.95(6)& 1.33(5)& \cellcolor{cyan}0.91(4)& 0.70(3)& 0.57(2) & 0.50(2) \\ \hline 
 0.03 & 1.502 & 1.148 & \cellcolor{green}2.93(9)& \cellcolor{yellow}2.00(8)& 1.37(5)& \cellcolor{cyan}1.04(4)& 0.86(3) & 0.75(4) \\ \hline 
 0.04 & 2.002 & 1.531 & \cellcolor{pink}3.91(12)& 2.67(9)& \cellcolor{yellow}1.82(7)& 1.39(5)& \cellcolor{cyan}1.14(5) & \cellcolor{cyan}1.00(5) \\ \hline 
 0.05 & 2.503 & 1.914 & 4.88(16)& \cellcolor{green}3.33(12)& 2.28(9)& 1.74(6)& 1.43(6) & 1.25(6) \\ \hline 
 0.06 & 3.004 & 2.296  & 5.86(19)& \cellcolor{pink}4.00(14)& 2.73(11)& \cellcolor{yellow}2.09(8) & 1.72(7) & 1.50(7)\\ \hline 
 0.07 & 3.504 & 2.679 & 6.8(2)& 4.67(16)& \cellcolor{green}3.19(13)& 2.43(9)& \cellcolor{yellow}2.00(8) & 1.75(8) \\ \hline 
 0.08 & 4.005 & 3.062 & 7.8(2)& 5.33(19)& 3.64(14)& 2.78(10)& 2.29(9) & \cellcolor{yellow}2.00(9)\\ \hline 
 0.09 & 4.505 & 3.445 & 8.8(3)& 6.0(2)& \cellcolor{pink}4.10(16)& \cellcolor{green}3.13(12)& 2.58(10) & 2.25(11) \\ \hline 
 0.10 & 5.006 & 3.827 & 9.8(3)& 6.7(2)& 4.55(18)& 3.48(13)& \cellcolor{green}2.86(11) & 2.50(12) \\ \hline 
 0.15 & 7.509 & 5.741 & x& x& 6.8(3)& 5.21(19)& \cellcolor{pink}4.29(17)& x\\ \hline
    \end{tabular}
    \caption{\textbf{Conversion table for} $\mathbf{\arga{\tilde{u}_\mathrm{max},\xi}}$.
    The random strains $\tilde{\gamma}_\mathrm{max}$ are straightforwardly obtained from Eq.~\eqref{eq:gammaTilde-sigmaTilde-Def}, since we kept the number of particles and the system size fixed in all our simulations.
    $\kappa_0(\xi)$ are measured, and the corresponding effective strains are computed. Parentheses show 95\% confidence intervals on the last digit (or digits), and values with $\sqrt{\kappa_0} \tilde{\gamma}_{\mathrm{max}} \approx 0.01$, $0.02$, $0.03$, and $0.04$ are highlighted in cyan, yellow, green, and pink respectively.
    }
    \label{tab:list-parameters}
\end{table}

\begin{figure}[htbp]
\begin{center}
\subfigure{\includegraphics[width=0.8\columnwidth]{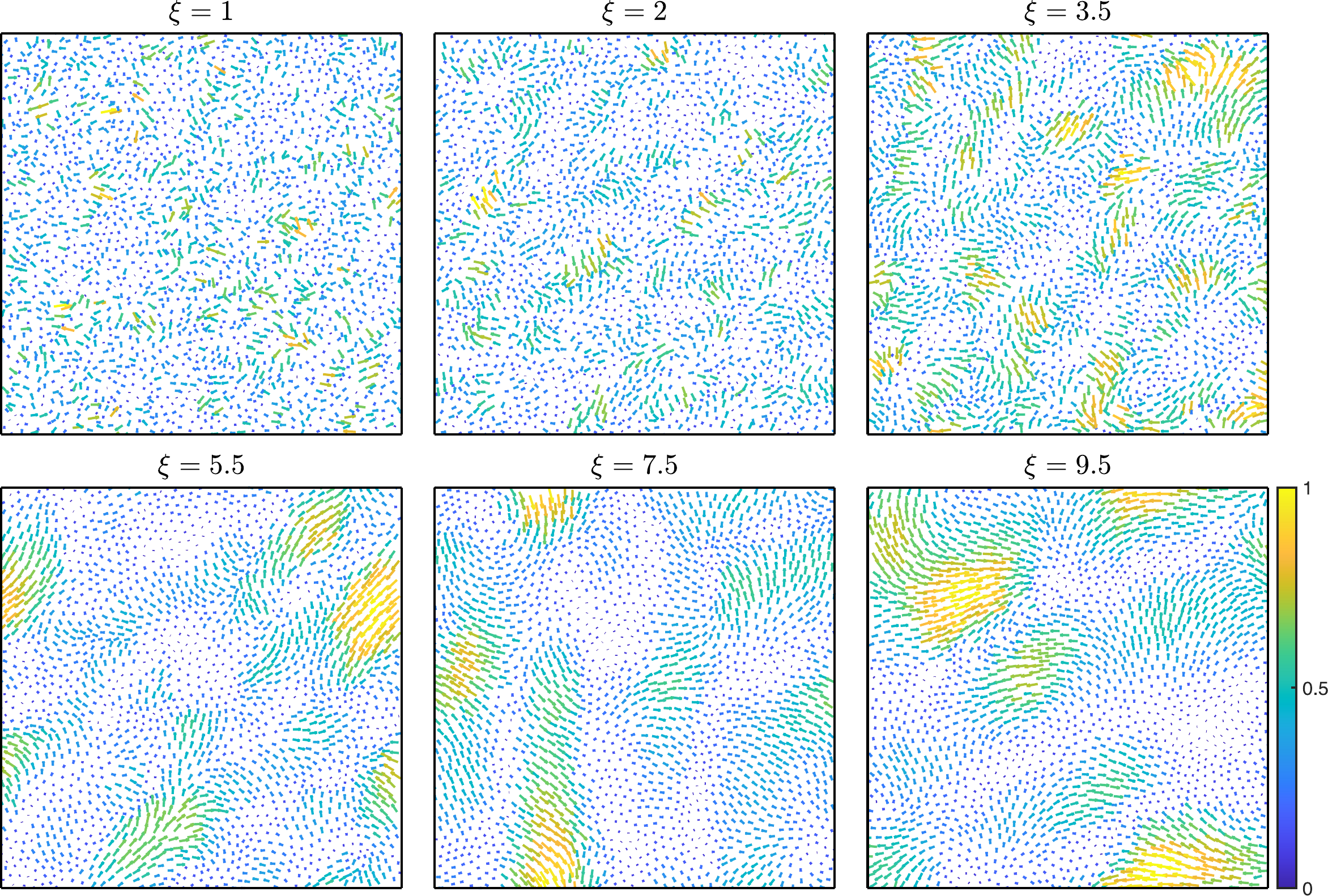}}
\caption{
\textbf{Examples of driving patterns for increasing correlation length $\xi$}. Distances are expressed in the units of the smallest particle diameter.
Note that vectors are \emph{not} unitary, as both the amplitude and orientation of the local displacements vectors are from a distribution.
}
\label{fig-snapshots-tuning-xi}
\end{center}
\end{figure}

\clearpage
\section{Individual and averaged stress-strain curves}
\label{sec-stress-strain-curves-compil}

\begin{figure}[htbp]
\begin{center}
\subfigure{\includegraphics[width=\columnwidth]{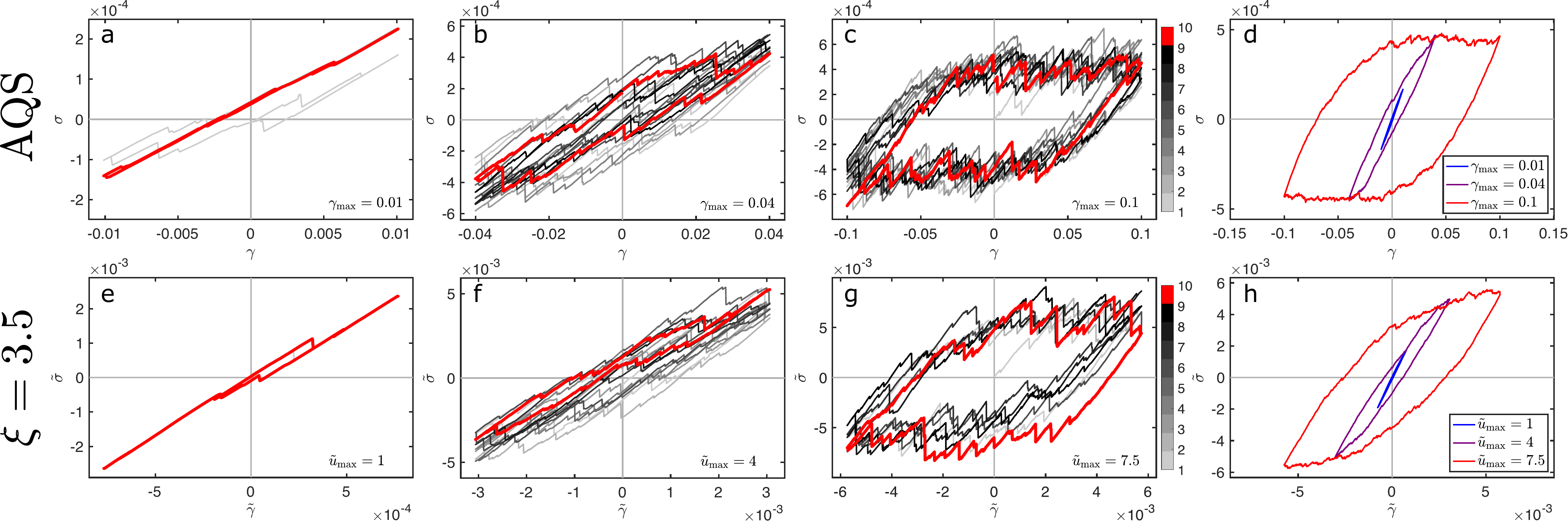}}
\caption{\textbf{Stress-strain curves plotted for a single configuration} driven by AQS at amplitudes (a) $\gamma_\mathrm{max}=0.01$, (b) $\gamma_\mathrm{max}=0.04$, and (c) $\gamma_\mathrm{max}=0.1$, and AQRD with $\xi=3.5$ at amplitudes (e) $\tilde{u}_\mathrm{max}=1$, (f) $\tilde{u}_\mathrm{max}=4$, and (g) $\tilde{u}_\mathrm{max}=7.5$.
Each cycle $t$ is plotted in a different shade of gray, as indicated in the color bar associated with (c) and (g) respectively.
The final cycle (which represents the limit cycle if it is achieved) is plotted in red.
The average curves taken over 30 configurations for each amplitude are given for (d)~AQS and (h)~AQRD.
Plots (d) and (h) are identical to the plots in Figs.~1b and~1d of the main text, and are repeated here for clarity.
}
\label{fig:SI-individual-trajectories}
\end{center}
\end{figure}

In Fig.~1 in the main text, we showed two examples of averaged limit cycles:
\begin{itemize}
\item for AQS at ${\gamma_\mathrm{max} \in \arga{0.01,0.04,0.1}}$,
\item for AQRD with ${\xi=3.5}$ at ${\tilde{u}_\mathrm{max}/L \in \arga{0.02,0.08,0.15}}$,
and after conversion 
$\sqrt{\kappa_0} \tilde{\gamma}_{\mathrm{max}}\in \arga{0.91,3.64,6.8} \times 10^{-2}$.
\end{itemize}
In Fig.~\ref{fig:SI-individual-trajectories}, we provide examples of the individual stress-strain curves upon cycling,
with a reprinting of the averaged curves.

\vline

\section{Driving pattern characterization vs average elastic modulus.}
\label{sec-effective-pattern-characterization-elastic-modulus-SI}

In Fig.~2 in the main text, we compare the driving patterns and the elastic modulus on the last cycle (${t=10}$).
More specifically in the main plot, we look at the ratio $\mathfrak{F}(\xi,\tilde{\gamma}_\mathrm{max})/\mathfrak{F}_0^\mathrm{AQS}$ as a function of the effective strain ${\sqrt{\kappa_0(\xi)} \tilde{\gamma}_{\mathrm{max}}}$,
where we chose ${\mathfrak{F}_0^{\mathrm{AQS}}/\ell^2=3.2\times10^{-6}}$ as a unique reference value independent of the subsequent driving (hence compatible with an agnostic analysis of the data).
It quantifies how much more disordered the driving pattern becomes upon iterative cycling.

\vline

In the inset, we compare four ratios as a function of $\xi$, which characterize either the disorder of the driving pattern, or the minimal statistical features of the landscape explored by the system (i.e., the typical \emph{apparent} elastic modulus):
\begin{equation}
\begin{split}
\text{[initial pattern]} \quad
F^r_0 (\xi) \equiv \frac{\mathfrak{F}_0(\xi)}{\mathfrak{F}_0^\mathrm{AQS}}
\, , \quad
& \text{[last-cycle pattern]} \quad
F^r_{\tilde{\gamma}_\mathrm{max}}(\xi) \equiv \frac{\mathfrak{F}(\xi,\tilde{\gamma}_\mathrm{max})}{\mathfrak{F}_0^\mathrm{AQS}(\gamma_\mathrm{max})} \, ,
\\
\text{[initial landscape]} \quad
\kappa_0(\xi)=\frac{\mu_0^{\mathrm{AQRD}}(\xi)}{\mu_0^\mathrm{AQS}}
\, , \quad
& \text{[last-cycle landscape]} \quad
\kappa(\xi,\tilde{\gamma}_\mathrm{max}))=\frac{\mu^{\mathrm{AQRD}}(\xi,\tilde{\gamma}_\mathrm{max}))}{\mu^\mathrm{AQS}(\gamma_\mathrm{max})} \, .
\end{split}
\end{equation}
Here we focused on
$\gamma_\mathrm{max}\approx \sqrt{\kappa_0(\xi)}\tilde{\gamma}_\mathrm{max} \approx 0.02$.
The driving pattern is characterized by a direct analysis of the displacement vectors ${\vert c \rangle=\arga{\mathbf{c}_i}}$ between interacting pairs of particles, and is defined as:
\begin{equation}
\frac{\mathfrak{F} (\xi)}{\ell^2}
\equiv d \, \text{Var} \argp{\vert  \mathbf{c}_i - \mathbf{c_j} \vert}
= \frac{d}{N_c} \sum_{\langle ij \rangle} \vert\vert \mathbf{c}_i - \mathbf{c_j} \vert\vert^2
\end{equation}
where $\langle ij \rangle$ denotes the $N_c$ interacting pairs and $\ell$ is the average distance between contacting particles.
Note that, since we are dealing with ratios of $\mathfrak{F}$, the specific value of $\ell$ is irrelevant.
The elastic modulus is obtained by linear response on a given configuration, however it could also be computed analytically for a given configuration by assuming a mechanically stable minimum and testing the linear response on its associated elastic branch (see the SI of Ref.~\cite{morse_roy_agoritsas_2021_PNAS118_e2019909118}):
\begin{equation}
\begin{split}
& \text{[configuration snapshot]} \quad
\arga{\mathbf{r}_{ij} \equiv \mathbf{r}_i - \mathbf{r}_j \equiv r_{ij} \, \hat{\mathbf{r}}_{ij}} \quad \text{with} \quad \hat{\mathbf{r}}_{ij}^2=1 ,
\\
& \text{[instantaneous driving pattern]} \quad
\arga{\mathbf{c}_{ij} \equiv \mathbf{r}_i - \mathbf{r}_j \equiv c_{ij} \, \hat{\mathbf{c}}_{ij}}  \quad \text{with} \quad \hat{\mathbf{c}}_{ij}^2=1 ,
\\
\\
& \mu_{\text{AQRD}} [\arga{\mathbf{r}_{ij}(t),\mathbf{c}_{ij}}]
 = \sum_{\langle ij \rangle} c_{ij}^2 \argc{v''(r_{ij}) (\hat{\mathbf{r}}_{ij} \cdot \hat{\mathbf{c}}_{ij})^2 + \frac{v'(r_{ij})}{r_{ij}} }
\\
& \quad \quad \quad \quad \quad \quad \quad \quad \stackrel{\text{(distr.)}}{\sim} \text{Var}(c_{ij}) \, ,
\end{split}
\end{equation}
%
where $v(r)$ is the radial pairwise interaction. We consider a Hertzian potential, but the definition is generic to a soft/hard-core potential.
The `(distr.)' denotes a scaling in distribution, meaning that the (random) distribution of apparent elastic modulii in a given configuration scales as the variance of the driving pattern; \emph{this explains why we expect a similar scaling between $\kappa$ and $\mathfrak{F}$}.
By considering ratios of these quantities with respect to its AQS values, we get rid of the overall finite-size factor. The remaining discrepancy then depends on the interaction potential.

Both $\kappa$ and $\mathfrak{F}$ are thus physical quantities measurable in an agnostic way from a given snapshot of the system.
In the inifinite-dimensional limit detailed in Ref.~\cite{agoritsas_2021_JStatMech2021_033501}, we would implicitly have ${\mathfrak{F}_0^{\mathrm{AQS}}/\ell^2=1}$ hence the analytical prediction ${\kappa_0 (\xi) = \mathfrak{F}_0(\xi)/\ell^2}$.
The correct way to use this prediction is instead
${\kappa_0 (\xi) \stackrel{(\mathrm{distr.})}{=} \frac{\mathfrak{F}_0(\xi)/\ell^2}{\mathfrak{F}_0^{\mathrm{AQS}}/\ell^2}=\mathfrak{F}_0(\xi)/\mathfrak{F}_0^{\mathrm{AQS}}}$,
though this does not change any of the conclusions presented in Ref.~\cite{morse_roy_agoritsas_2021_PNAS118_e2019909118}.
At last, the variance of the driving pattern is useful to rationalize the scaling in $\xi$ of the mechanical response features,
but the typical elastic modulus is physically more meaningful as it also includes information from the actual configuration of the system.

\begin{figure}[htbp]
\begin{center}
\subfigure{\includegraphics[width=0.9\columnwidth]{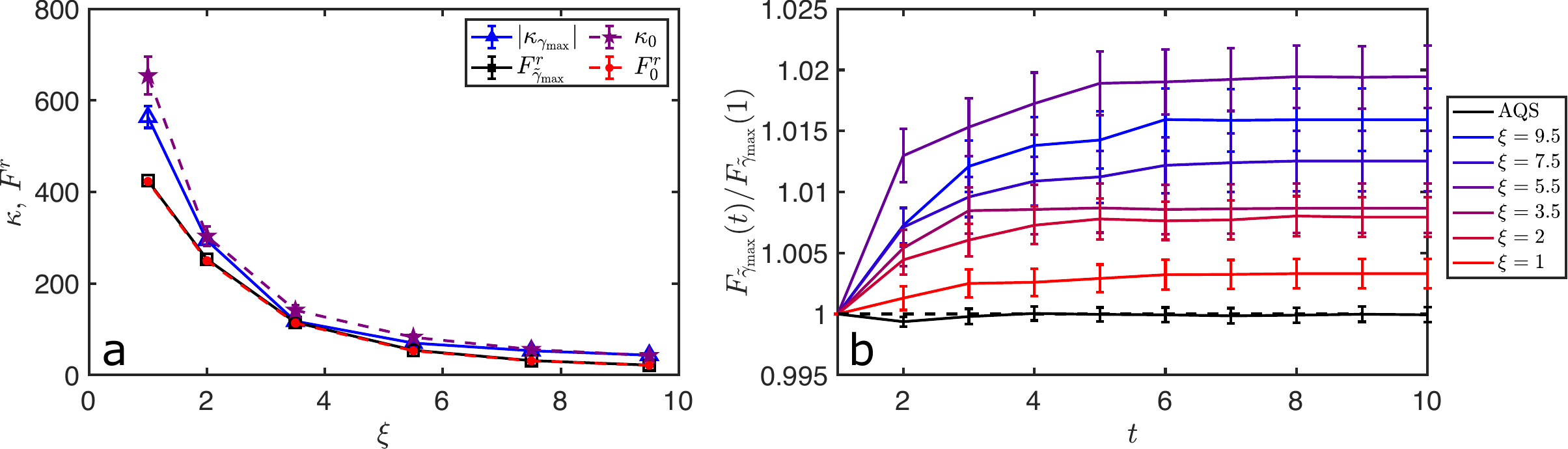}}
\caption{
\textbf{Measuring the disorder of the driving field and its relative lack of evolution upon cycling.}
(a)~The same data in the inset of Fig.~2 from the main text is plotted on a linear scale with the error-bars associated with taking an average over the 30 initial configurations.
These are compatible with power-law exponents of $-1.4(3)$ for $F^r_0$, $-1.23(5)$ for $\kappa_0$, $-1.4(3)$ for $F^r_{\tilde{\gamma}_\mathrm{max}}$, and $-1.19(16)$ for $\kappa_{\tilde{\gamma}_\mathrm{max}}$, with error bars indicating $95\%$ confidence intervals, consistent with \cite{morse_roy_agoritsas_2021_PNAS118_e2019909118,agoritsas_2021_JStatMech2021_033501}.
(b)~The driving pattern is taken from snapshots at the maximum strain in systems where $\gamma_\mathrm{max},\,\sqrt{\kappa_0}\tilde{\gamma}_\mathrm{max}\approx 0.02$ as a function of the number of cycles. The relative change in the field and the plateau upon cycling indicates that the scaling collapse from the initial $\kappa_0$ should still hold.
}
\label{fig:SI-memory-compil}
\end{center}
\end{figure}

For completeness, we report in Fig.~\ref{fig:SI-memory-compil} an alternative representation of the data plotted in Fig.~2 of the main text.

\section{Geometrical features of hysteretic limit cycles}
\label{sec-geometrical-features-hysteresis}

In Fig.~3 in the main text,
we report the collapse of the average limit cycles, or at least at the last cycle (${t=10}$), with the rescaling factor ${\kappa_0(\xi)}$.
The notion of limit cycle itself is better characterized via the MSD plots as in Fig.~4(a,b) in the main text.

Note that, because we the same range of strain amplitudes $\gamma_\mathrm{max}$ in AQS and displacement amplitudes $\tilde{u}_\mathrm{max}$ in AQRD, upon rescaling with $\sqrt{\kappa}$ our simulation range is effectively compressed when we increase $\xi$ (red to blue).
Thus in order to examine the quality of the collapse, we have to compare points which have a comparable effective strain \emph{after} rescaling with $\kappa_0$.
A small mismatch between curves is thus expected, as \emph{we do not work exactly at the same effective strains}.
The situation was much simpler in Ref.~\cite{morse_roy_agoritsas_2021_PNAS118_e2019909118}, where we collapsed data sets in the very first cycle at comparable effective strains
$\arga{\gamma,\tilde{\gamma}}$, so densifying the datapoints could be achieved simply by tuning the small increments $\arga{\Delta \gamma,\Delta \tilde{u}}$.
Here we instead compare maximum strain/displacement amplitudes with runs over ${t=10}$ cycles, which renders the addition of new datapoints much more tedious, but at the same time much more informative.

\section{Quantifying amnesia}
\label{sec-quantifying-amnesia}

\begin{figure*}[htp]
\begin{center}
\includegraphics[width=0.9\linewidth]{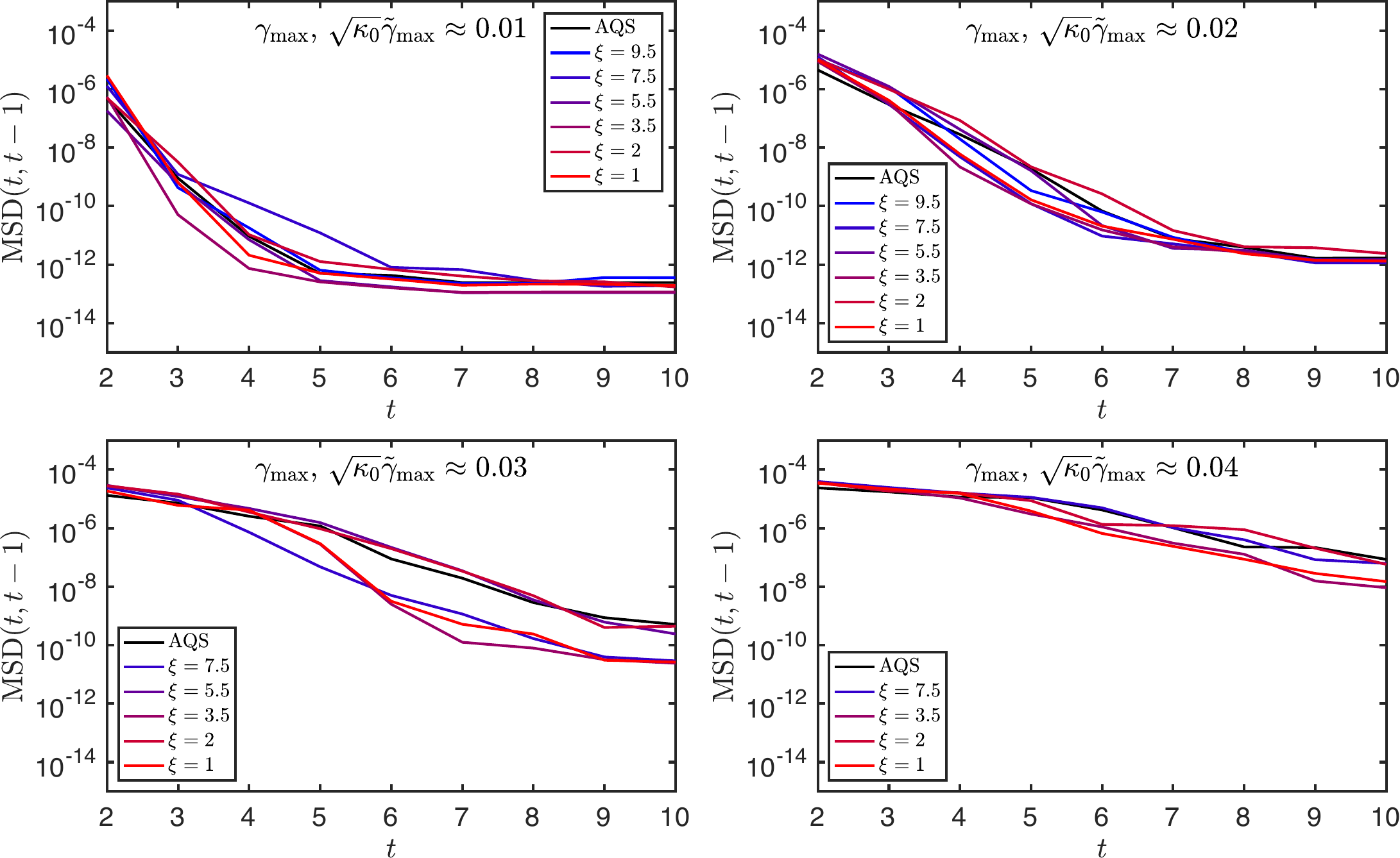}
\caption{\textbf{Collapse of the amnesia rate upon rescaling of the random strain.}
See Table~\ref{tab:list-parameters} for the corresponding composite datasets.
While the collapse for $0.03$ captures the correct trend, the spread in behavior stems from the wide range of effective strains available for the corresponding composite AQRD dataset, namely $\sqrt{\kappa_0}\tilde{\gamma}_\mathrm{max}=\arga{2.93, 3.33, 3.19, 3.13, 2.86} \times 10^{-2}$. All collapses are expected to improve with greater precision in the choice of $\sqrt{\kappa_0}\tilde{\gamma}_\mathrm{max}$ and with further averaging.}
\label{fig-memory-SI-bis}
\end{center}
\end{figure*}

In Fig.~4(b) in the main text, we show the collapse of the average MSD functions for a given effective strain ${\sqrt{\kappa_0} \tilde{\gamma}_\mathrm{max} \approx 0.02}$.
In Fig.~\ref{fig-memory-SI-bis}, we plot the average MSD for additional values of effective strains in the `pre-yielding' regime at ${\sqrt{\kappa_0} \tilde{\gamma}_\mathrm{max} \approx 0.01, 0.02, 0.03}$ and $0.04$, showing that the curves collapse in each case, which indicates that the behavior is generic.

\begin{figure*}[htp]
\begin{center}
\includegraphics[width=\linewidth]{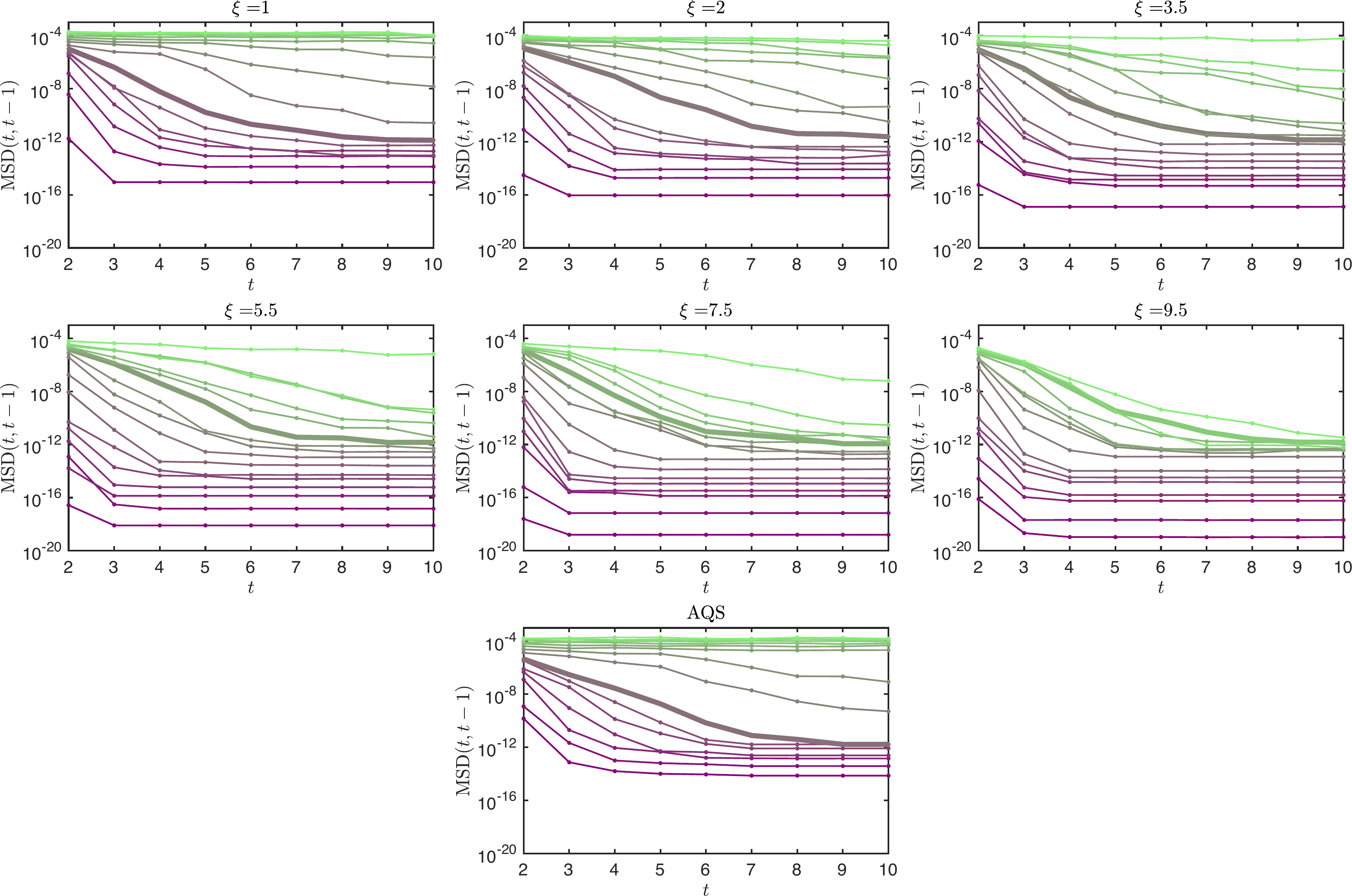}
\caption{\textbf{Quantifying amnesia upon cycling}. MSD function at increased spatial correlation $\xi$, averaged over 30 initial configurations, for increasing maximum strain. The bold line in each indicates ${\sqrt{\kappa_0}\tilde{\gamma}_{\mathrm{max}} \approx 0.02}$.
The color scale linearly interpolates between the smallest value of $\tilde{u}_\mathrm{max}$ (purple) to the largest value of $\tilde{u}_\mathrm{max}$ (green) for a given value of $\xi$ (similar for AQS, substituting $\gamma_\mathrm{max}$).
See Table~\ref{tab:list-parameters} for the list of $\tilde{u}_\mathrm{max}$ used in each system. 
}
\label{fig-memory-SI}
\end{center}
\end{figure*}

In Fig.~4(a) in the main text, we provide one example of the MSD function at fixed spatial correlation ${\xi=2}$, averaged over the 30 initial conditions, for increasing $\tilde{u}_\mathrm{max}$.
In Fig.~\ref{fig-memory-SI} we complete this picture by providing the corresponding plots for AQRD at increasing $\xi$ and for AQS, including all of the available datapoints.

\clearpage

\end{document}